\documentclass[preprint,aps,prb]{revtex4}
\usepackage{graphicx}
\usepackage{dcolumn}
\usepackage{bm}

\usepackage{titlesec}
\titleformat*{\section}{\flushleft \bf \large}
\titleformat*{\subsection}{\flushleft \bf}
\titleformat*{\subsubsection}{\flushleft}
\begin{document}

\title{Activated molecular adsorption of CO on the Be (0001) surface: A density-functional theory study}

\author{Shuang-Xi Wang$^{1,2}$, Yu Yang$^{2}$, Bo Sun$^2$, Rong-Wu Li$^1$, Shao-Jun Liu$^1$, Ping Zhang$^2$\footnote{Corresponding
author. E-mail address: zhang\underline{ }ping@iapcm.ac.cn}}

\affiliation{$^1$Department of Physics, Beijing Normal University,
Beijing 100875, China}

\affiliation{$^2$LCP, Institute of Applied Physics and Computational
Mathematics, P.O. Box 8009, Beijing 100088, People's Republic of
China}

\begin{abstract}
Using first-principles calculations, we systematically study the
adsorption behaviors of molecular CO on the Be (0001) surface. By
calculating the potential energy surfaces, we find that CO
molecularly adsorbs on the Be surface with small energy barriers.
The most stable adsorption state is found to be the one at the
surface fcc hollow site, and the one at the surface top site is the
adsorption state that has the smallest energy barrier. Based on
electronic structure analysis, we further reveal that during the
molecular adsorption, the $5\sigma$ bonding and $2\pi$ antibonding
orbitals of CO hybridize with $s$ and $p_z$ electronic states of Be,
causing electrons to transfer from CO to Be.
\end{abstract}

\maketitle

\section{Introduction}

It is of great importance to study the adsorption and dissociation
of diatomic molecules on metal surfaces to meet many intrinsic
academic interests \cite{Darling1995,King1988}. Besides, the
adsorption of carbon monoxide (CO) on metal surfaces is one
important part in many technological processes such as heterogeneous
catalysis \cite{Nilsson2004}, corrosion
\cite{Heusler1961,Marcus2002}, gas sensing \cite{Srinivasan1998} and
exhaust gas removing \cite{Libuda2005}, and so has been studied
extensively. For the adsorption on transition metal surfaces, it has
been found that CO molecularly adsorbs on almost all transition
metal surfaces without any energy barriers
\cite{Hammer1996,Hellman2005}. In comparison with the vast studies
on the adsorption properties of CO on transition metals, the
adsorption of CO on simple metal surfaces has seldom been studied
yet because of their irrelevance to heterogeneous catalysis. Only
very recently, Hellman {\it et al.} predicted by using first
principles calculations that CO adsorbs on the Al (111) surface
molecularly with a small energy barrier \cite{Hellman2005},
different from the adsorption behaviors of CO on transition metal
surfaces.

However, many studies are now being applied to explore the
possibilities to use fabricated simple metal structures instead of
transition metals in catalytical processes \cite{Ma2007}, because
specially fabricated metal structures have displayed quite different
catalytic properties with bulk metal materials \cite{Chen2004}, and
simple metals are always lighter and cheaper than transition metals.
So the adsorption properties of CO on simple metals also need to be
studied, specially considering that so scarce theoretical studies of
this issue are available in literature and it is impossible for one
to give any concluding remarks on the common nature of the CO
adsorption on simple $sp$ metal surfaces. Motivated by this
observation, in this paper we use first-principles calculations to
systematically study the adsorption of CO on the Be (0001) surface.
The reason why we choose Be as the prototype for simple metals is
that Be has long been used in nuclear reactors as air cleaners to
adsorb many kinds of residue gases \cite{Zalkind1997,Argentina2000,
Zalkind2002}, and Be is the second lightest metal. Besides and
prominently, differing from the bulk, Be surfaces have a large
directional $s$ and $p$ electronic distributions around the Fermi
energy \cite{Boettger1986} and thus may display unique interaction
with diatomic molecules \cite{Zhang2009}.

By calculating the adiabatic potential-energy surfaces (PESs) and
analyzing the projected density of states (PDOS), we obtain the
adsorption properties of CO on the Be (0001) surface. It is found
that CO adsorbs on the Be surface molecularly with small energy
barriers. The adsorption interactions are mainly contributed by
electrons transfer from bonding orbitals of CO to $s$ and $p_z$
electronic states of Be (0001) surface layer, as well as by strong
electronic hybridization between antibonding orbital of CO and $p_x$
states of Be.

\section{Calculation method}

Our calculations were performed within density functional theory
using the Vienna \textit{ab-initio} simulation package (VASP)
\cite{VASP}. The PW91 \cite{PW91} generalized gradient approximation
and the projector-augmented wave potential \cite{PAW} were employed
to describe the exchange-correlation energy and the electron-ion
interaction, respectively. The cutoff energy for the plane wave
expansion was set to 520 eV. The Be (0001) surface was modeled by a
slab composing of five atomic layers and a vacuum region of 20 \AA.
A 2 $\times$ 2 supercell, in which each monolayer contains four Be
atoms, was adopted in the study of the CO adsorption since our test
calculations have showed that it is sufficiently large to avoid the
interaction between adjacent CO molecules. Integration over the
Brillouin zone was done using the Monkhorst-Pack scheme
\cite{Monkhorst1976} with 11 $\times$ 11 $\times$ 1 grid points. And
a Fermi broadening \cite{Weinert1992} of 0.1 eV was chosen to smear
the occupation of the bands around the Fermi energy ($E_{F}$) by a
finite-$T$ Fermi function and extrapolating to $T$ = 0 K. The
calculation of the potential energy surface was interpolated to 121
points with different bond length ($d_{\mathrm{CO}}$) and height
($h$) of CO at each surface site. The calculated lattice constant of
bulk Be ($a$, $c$) and the bond length of a free CO molecule are
2.26 \AA, 3.58 \AA~ and 1.14 \AA, respectively, in good agreement
with the experimental values of 2.29 \AA, 3.588 \AA~
\cite{Wachowicz2001} and 1.13 \AA~ \cite{Huber1979}.

\section{Results and discussion}

Firstly, we do a geometry optimization for the clean Be (0001)
surface, during which the bottom two atomic layers of the Be surface
is fixed and other Be atoms are free to relax until the forces on
them are less than 0.02 eV/\AA~ along the $x$, $y$ and $z$
directions (as shown in Fig. 1). Then we calculate a series of
two-dimensional (2D) PES cuts for a CO molecule on the relaxed Be
(0001) surface. As depicted in Fig. 1, there are four high symmetry
sites on the Be (0001) surface, respectively the top, bridge (bri),
hcp and fcc hollow sites. Different from the studies on the
adsorption of symmetrical diatomic molecules \cite{Yang2008}, a CO
molecule at each surface site has five different high-symmetry
orientations. And we here use $y1$ and $y2$, $z1$ and $z2$ to
differentiate C end-on and O end-on orientations. As shown in Fig.
1, there are in total 20 different adsorption channels for CO
molecule on the Be (0001) surface. By employing the notations in
Fig. 1, we will represent the channels as top-$x,y1,y2,z1,z2$,
bri-$x,y1,y2,z1,z2$, hcp-$x,y1,y2,z1,z2$ and fcc-$x,y1,y2,z1,z2$.

The calculated 2D PES cuts along the top-$z1$, top-$z2$, and top-$x$
channels are shown in Figs. 2(a)-(c) respectively, which are quite
different from each other. The calculated 2D PES cuts along other
adsorption channels at the top site have similar shapes with that
along the top-$x$ channel and thus are not plotted here for brevity.
As shown in Fig. 2, we find that the adsorption of molecular CO at
the surface top site only occurs when the CO molecule is
perpendicular to the metal surface with C end-on orientation.
Similar results have been obtained in the CO/Al(111) system
\cite{Hammer1996,Hellman2005}. In following discussion the
adsorption state for molecular CO along the top-$z1$ channel will be
called as the TZ state. In the PES cut along the top-$z2$ channel,
no CO molecular adsorption states exist, and the energy needed to
separate the C and O atoms by 1.80 \AA~ is very high ($\sim$ 6.8
eV). For the adsorption process along the top-$x$ channel and all
similar (namely, top-$y1$ and top-$y2$) channels, there are no
adsorption states either. However, the energy needed to separate the
C and O atoms by 1.8 \AA~ is relatively small ($\sim$ 3.05 eV). More
interestingly, we find that for a CO molecule to reach the TZ state,
it needs to overcome an energy barrier of 0.11 eV, which is quite
different from the molecular adsorption of CO on transition metal
surfaces, where the energy barrier is zero.

Through systematic PES calculations, we find that there are always
molecular adsorption states when the CO molecule is perpendicular to
the Be(0001) surface with C end-on orientations. For illustration
the PES cuts along the hcp-, fcc-, and bri-$z1$ channels are shown
in Figs. 3(a)-(c), with the corresponding molecular adsorption
states named as the HZ, FZ, and BZ states, respectively. Figure 3(d)
summarizes the four minimum-energy paths for the molecular
adsorption of CO along the top-, hcp-, fcc- and bri-$z1$ channels.
The corresponding energy barriers for the transition to the
molecular adsorption states along these four paths are respectively
0.11, 0.32, 0.30 and 0.28 eV. Thus, on one hand it clearly shows
that the energy barrier is the smallest for CO to evolve into the TZ
state. On the other hand, the energy barriers for these four
molecularly adsorbed states are all relatively small, indicating
that in experiment they should all be observed at room temperature.

In addition to the four molecularly adsorbed states revealed in Fig.
3(d), we also find another kind of local-minimum states, which is
along the bri-$y1$ and -$y2$ channels. Figure 4 and its inset
respectively show the PES cut along the bri-$y1$ channel and the
corresponding minimum energy path (the situation for the bri-$y2$
channel is quite similar and so is not shown here). However, one can
see that the energy barrier to evolve into this local-minimum state
(named as LM1) is as large as 1.35 eV. Moreover, the total energy of
the LM1 state is 0.80 eV higher than that in the free case (namely,
clean Be surface plus isolated CO molecule), indicating that this
kind of molecularly adsorbed state seldom happen in practice.

Now let us see the stability of these molecularly adsorbed states by
calculating their adsorption energies as follows
\begin{equation}\label{Ead}
E_{\rm ad}=E_{\rm CO}+E_{\rm Be(0001)}-E_{\rm CO/Be(0001)},
\end{equation}
where $E_{\rm CO}$, $E_{\rm Be(0001)}$, and $E_{\rm CO/Be(0001)}$
are the total energies of the CO molecule, the clean Be surface, and
the adsorption system respectively. According to this definition, a
positive value of $E_{ad}$ indicates that the adsorption is
exothermic (stable) with respect to a free CO molecule and a
negative value indicates endothermic (unstable) reaction. The
calculated adsorption energies are 0.64, 0.61, 0.68 and 0.51 eV for
the TZ, HZ, FZ and BZ states, and -0.80 and -0.91 eV for the two
local-minimum states along the bri-$y1$ and
$y2$ channels, respectively. 
From the calculated adsorption energies, we can see that the FZ
state is a little more stable than the TZ, HZ and BZ states. In
consideration of our previous result that there are less electrons
distributed at the surface fcc hollow site than at other sites
\cite{Wang2009}, the relative stability for the four adsorption
states seems to indicate that the interactions between the CO
molecule and the Be(0001) surface are dominated by charge transfer
from CO to Be, which will be discussed in the follows.

As mentioned above, the energy differences among the TZ, HZ, FZ and
BZ states are very tiny, and the energy barriers to reach them are
also small enough to be overcome at room temperatures. So for
further illustration, we here calculate the PDOS for all of these
molecularly adsorbed states. We find that the four adsorption states
have similar PDOS characters. As a typical example, we plot in Fig.
5 the PDOS for the CO molecule and one surface Be atom in the TZ
state. For comparison, the PDOS for the free CO molecule and clean
Be(0001) surface are also shown in Fig. 5. One can clearly see from
Fig. 5(a) that after adsorption, the $3\sigma$ bonding orbital of
the CO molecule negligibly changes, while the $4\sigma$ and $1\pi$
bonding orbitals change a little by losing few electrons, and the
$5\sigma$ bonding and $2\pi$ antibonding orbitals almost disappear.
So during the molecule-metal interaction, the electrons in the
$5\sigma$ bonding orbital of CO all transfer to the electronic
states of Be(0001) surface. Besides, we see that no charge transfer
from electronic states of Be back to the CO molecule as there are no
new orbital occupation in the PDOS of CO after adsorption. So the
net charge transfer during the adsorption is from CO to Be. This
result confirms our earlier speculation that the FZ state is stabler
than the TZ, HZ and BZ states because electrons are easier to
transfer from CO to Be at the surface fcc site. Moreover, since the
$5\sigma$ bonding orbital of CO is asymmetric in that more electrons
distribute around C than around O atom \cite{Fohlisch2000}, the
present PDOS analysis can also explain why only along the channels
with C end-on orientations there are molecularly adsorbed states.
The reason is that charge transfer happens easier when C atoms are
close enough to the Be surface.

From Fig. 5(b) and 5(c) one can see that the $s$ and $p_z$
electronic states of surface Be shift to lower energies during the
adsorption, indicating that they accept electrons from bonding
molecular orbitals of CO. In addition, there are two new peaks for
Be $s$ and $p_z$ states, aligning in energy with CO $4\sigma$ and
$1\pi$ orbitals, respectively, indicating as well an observable
covalency among these orbitals. Another fact is that at the energy
range above the Fermi energy, the $s$ and $p_z$ electronic states of
Be also change a lot by hybridizing with the $2\pi$ antibonding
orbital of CO. In total, the electronic states of Be that interact
with molecular orbitals of CO are mainly the $s$ and $p_z$ states.
And correspondingly, one can see from Figs. 5(b) and 5(c) that the
Be $p_x$/$p_y$ states change very little during the molecular
adsorption of CO.

To gain more insight into the nature of chemical bonding between CO
and Be during surface adsorption, we also analyze the difference
electron density $\Delta \rho(\mathbf{r})$ for the four stable
molecularly adsorbed states. Here $\Delta \rho(\mathbf{r})$ is
obtained by subtracting the electron densities of noninteracting
component systems,
$\rho^{\text{Be(0001)}}(\mathbf{r})+\rho^{\text{CO}}(\mathbf{r})$,
from the density $\rho(\mathbf{r})$ of the CO/Be(0001) system, while
retaining the atomic positions of the component systems at the same
location as in CO/Be(0001). As a typical example, the calculated
$\Delta \rho(\mathbf{r})$ for the TZ state is shown in Fig. 6. We
can see that there is a large charge depletion area between the C
and O atoms, showing the depopulation of the $5\sigma$ bonding
orbital of CO by transferring electrons out. And correspondingly,
there is a large charge accumulation near the Be (0001) surface,
indicating acceptance of electrons of the $s$ and $p_z$ electronic
states of the surface Be atoms. Remarkably, this charge accumulation
is highly directional instead of isotropically itinerant around the
surface. In this aspect, the acceptance of electron makes the
Be(0001) surface rather insulating than metallic. The same
conclusion can also be derived from Fig. 5, which shows that the
surface Be PDOS at the Fermi energy is largely decreased after
molecular adsorption of CO.

\section{Conclusion}

In conclusion, we have studied the adsorption behavior of CO on the
Be (0001) surface by using first-principles DFT method. Through
systematic PES calculations, we have revealed that the adsorption of
CO on the Be (0001) surface is molecular with small energy barriers.
The stable adsorption states all have C end-on orientations. The
most stable adsorption state for molecular CO has been found to be
along the fcc-$z1$ channel, while the adsorption state with the
smallest energy barrier is along the top-$z1$ channel. It has also
been shown that the interactions between CO and the Be surface are
dominated by charge transfer from CO to Be. We expect that the
present results are greatly helpful for the practical usage of Be
surfaces in adsorbing residual gases in experimental nuclear fusion
reactors.

\clearpage

\noindent\textbf{List of captions} \\

\noindent\textbf{Fig.1}~~~ (Color online). (a) The p($2\times2$)
surface cell of Be (0001) and four on-surface adsorption sites. Here
only the outmost two layers of the surface are shown. (b) The sketch
map showing that the molecule (with vertical or parallel
orientation) is initially away from the surface with a hight $h$.
Black, red and grey balls respectively represent for C, O and Be
atoms. \\

\noindent\textbf{Fig.2}~~~ (Color online). The 2D PES cuts for the
adsorption of CO along the (a) top-$x$, (b) $z1$, and (c) $z2$
channels. Here, the dashed lines denote the minimum-energy paths and
the energy intervals of contours are all 0.20 eV. \\

\noindent\textbf{Fig.3}~~~ (Color online). The 2D PES cuts for the
adsorption of CO along the (a) hcp-$z1$, (b) fcc-$z1$, and (c)
bri-$z1$ channels. The minimum-energy adsorption paths along the
top-, hcp-, fcc- and bri-$z1$ channels are summarized in (d). \\

\noindent\textbf{Fig.4}~~~  (Color online). The 2D PES cut for the
adsorption of CO along the bri-$y1$ channel, with the energy
interval of 0.20 eV. The inset shows the minimum-energy path along
this channel. \\

\noindent\textbf{Fig.5}~~~ (Color online). (a) PDOS for the CO
molecule before and after the adsorption. (b) and (c) PDOS for a
surface Be atom before and after the adsorption. The Fermi energy is
set to zero. \\

\noindent\textbf{Fig.6}~~~ (Color online) Difference electron
density for the most stable TZ adsorption state of CO on the Be
(0001) surface. Black, red, and grey balls denote C, O, and Be
atoms, respectively. Solid and dotted lines represent charge
accumulation and depletion, respectively.

\clearpage

\begin{figure}
\includegraphics[width=1.0\textwidth]{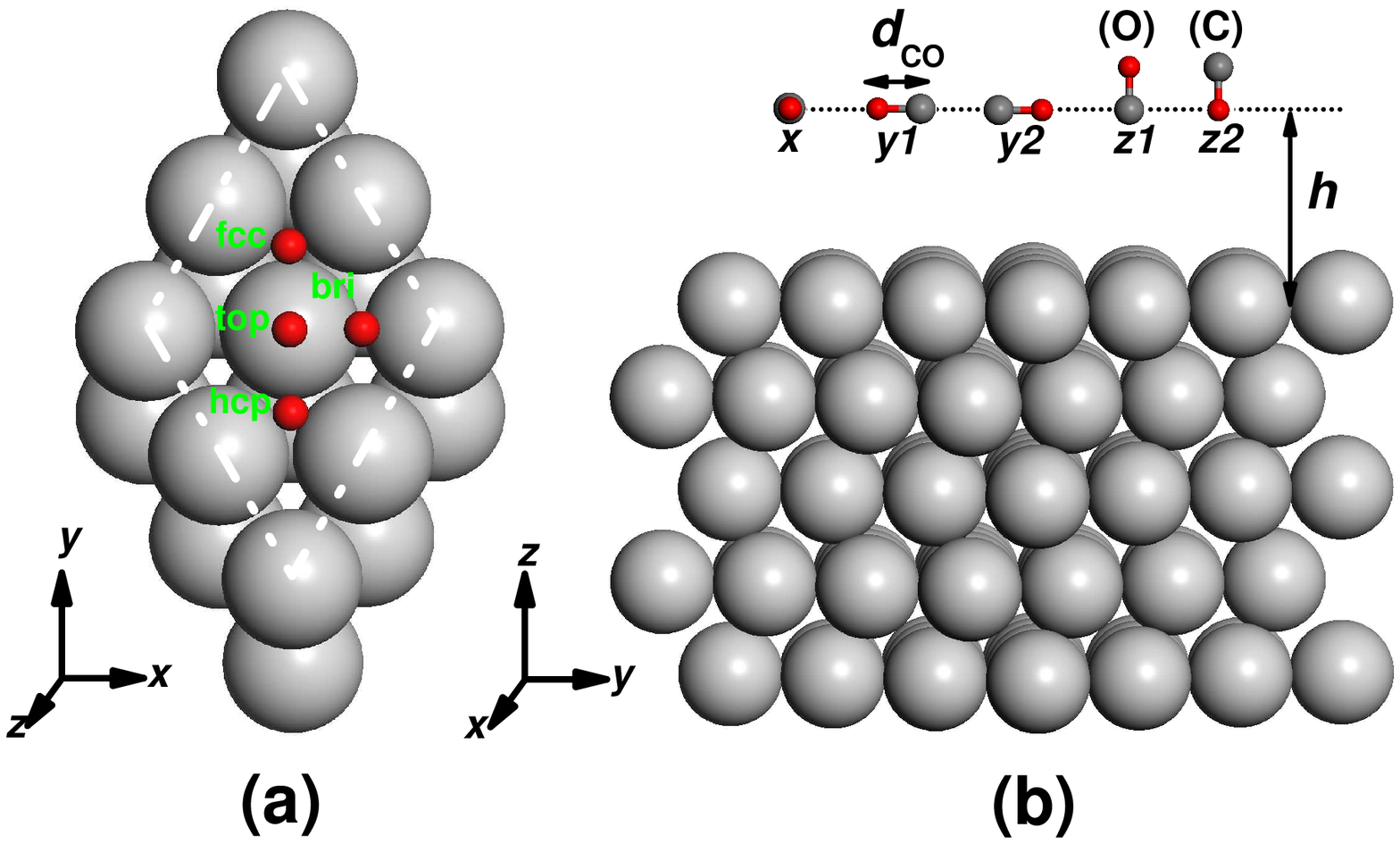}
\caption{\label{fig:fig1}}
\end{figure}
\clearpage
\begin{figure}
\includegraphics[width=1.0\textwidth]{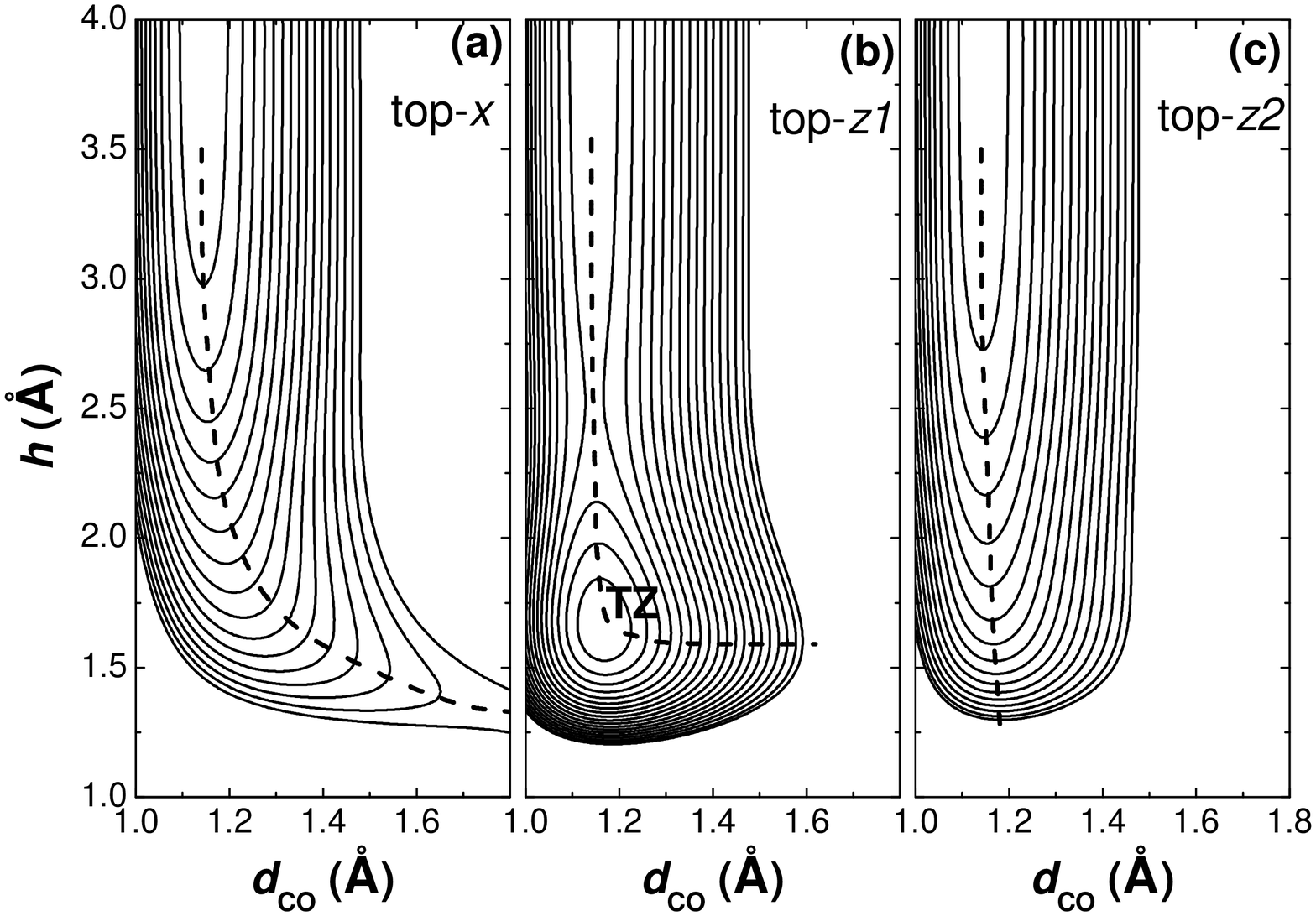}
\caption{\label{fig:fig2}}
\end{figure}
\clearpage
\begin{figure}
\includegraphics[width=1.0\textwidth]{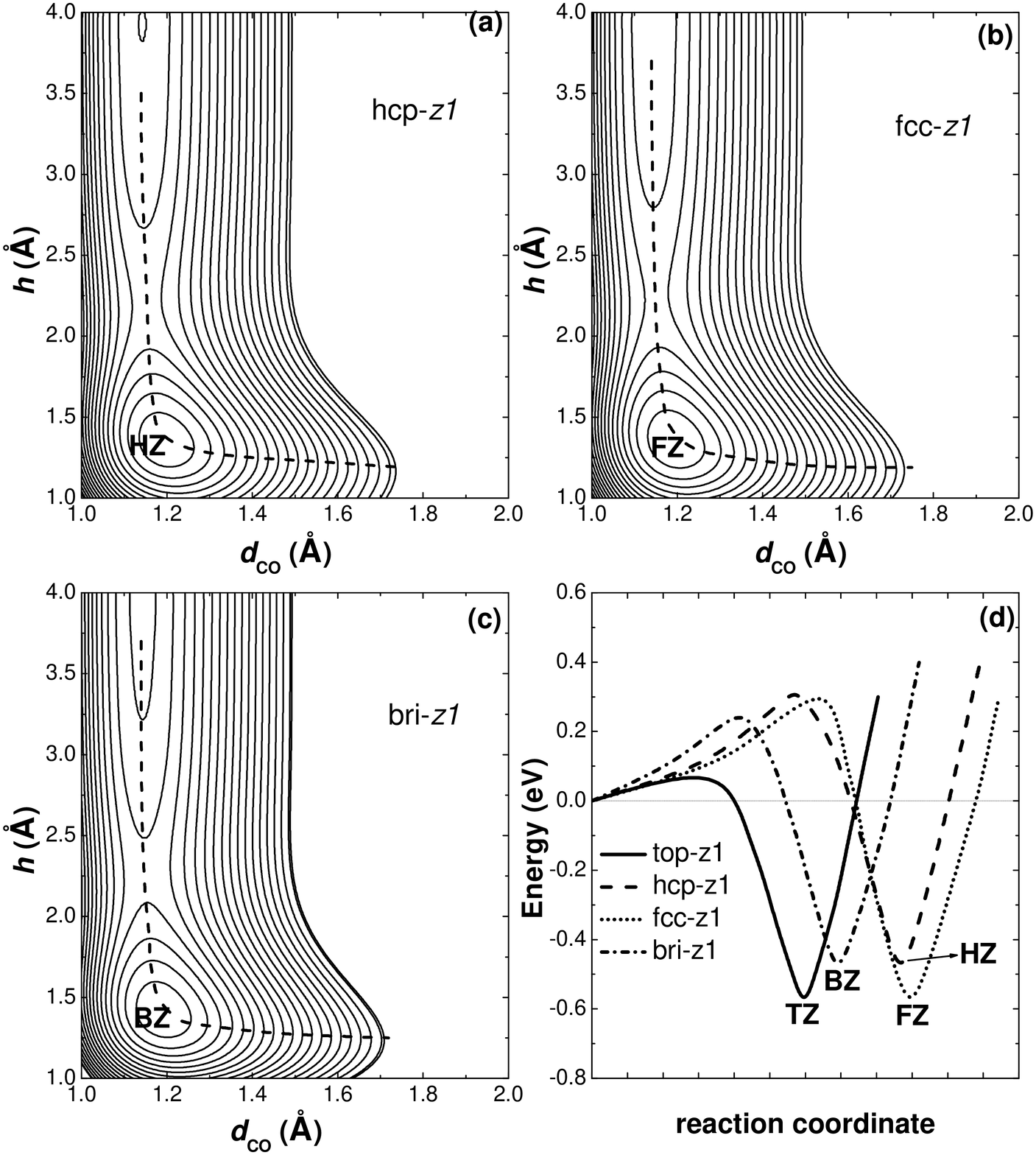}
\caption{\label{fig:fig3}}
\end{figure}
\clearpage
\begin{figure}
\includegraphics[width=1.0\textwidth]{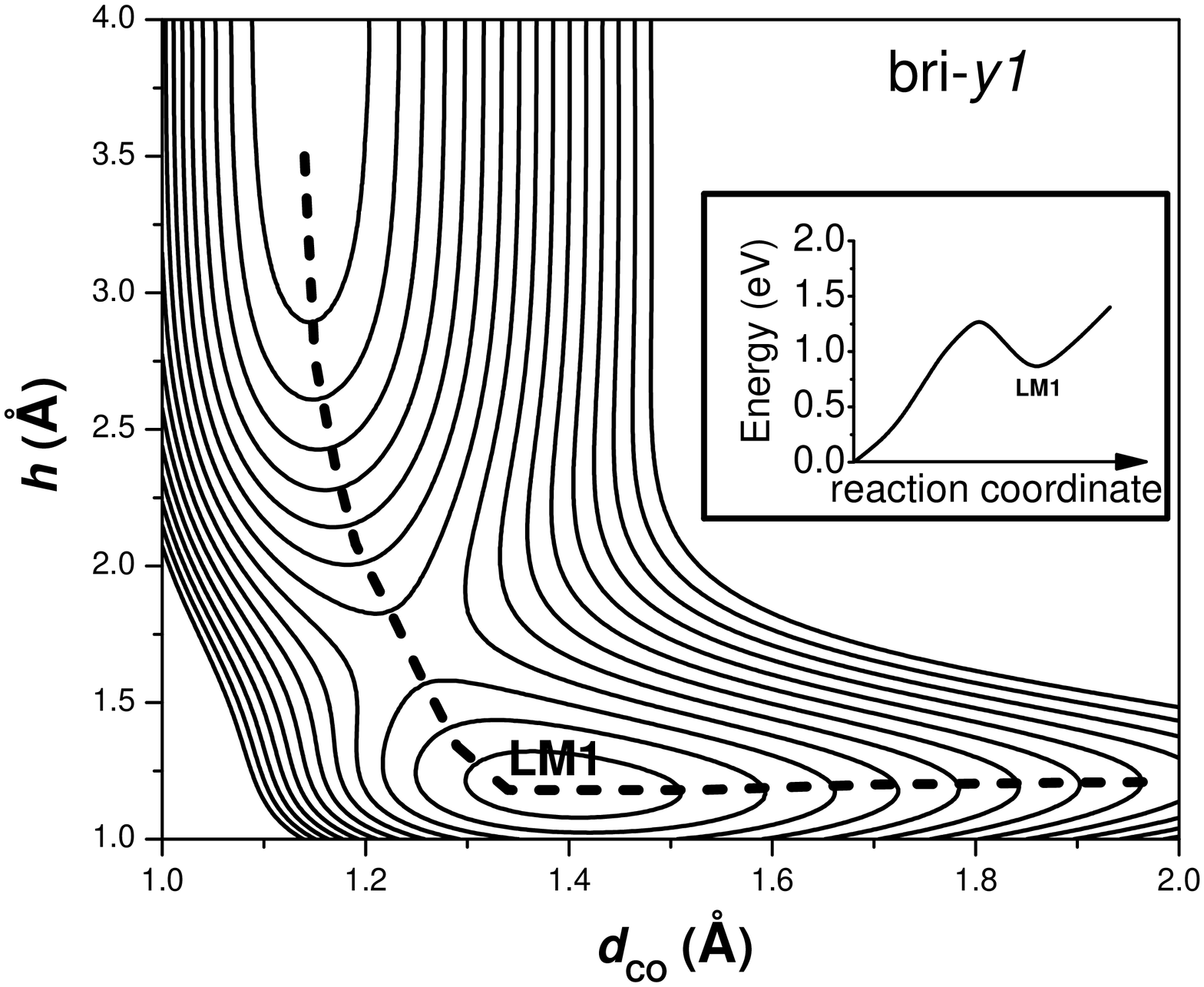}
\caption{\label{fig:fig4}}
\end{figure}
\clearpage
\begin{figure}
\includegraphics[width=1.0\textwidth]{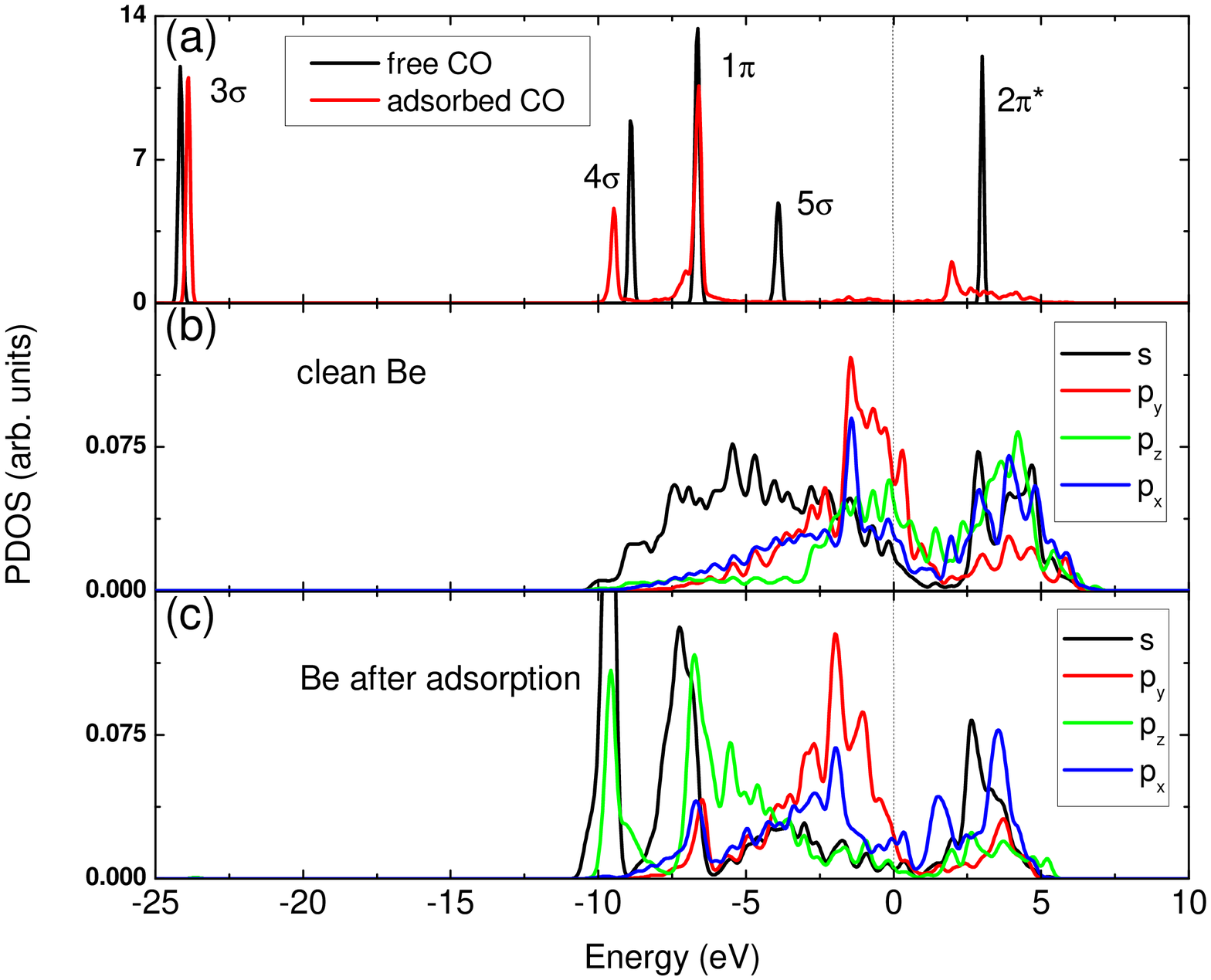}
\caption{\label{fig:fig5}}
\end{figure}
\clearpage
\begin{figure}
\includegraphics[width=1.0\textwidth]{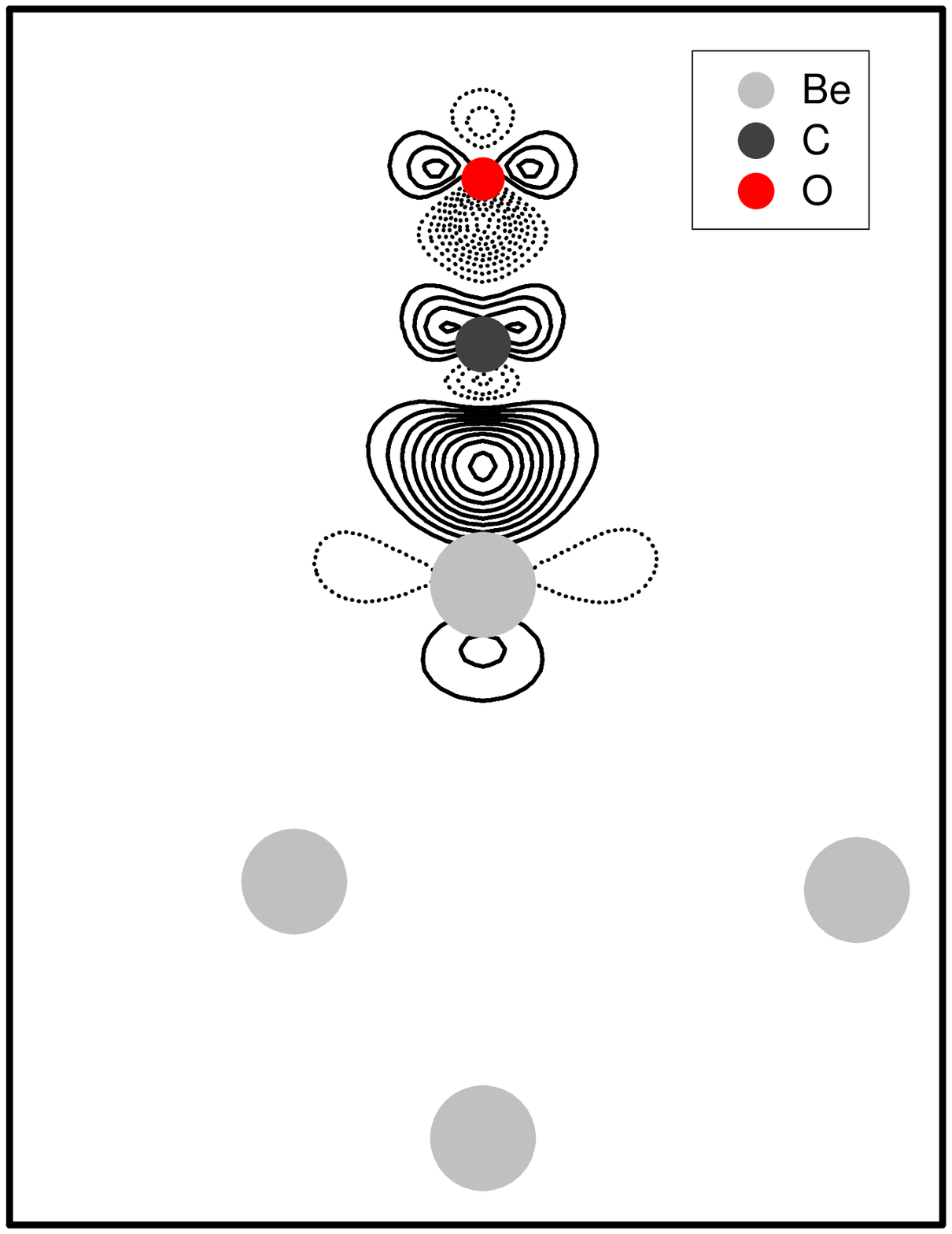}
\caption{\label{fig:fig6}}
\end{figure}
\end{document}